\documentclass[prb,twocolumn]{revtex4}
\usepackage{amsmath}
\usepackage{graphicx}
\usepackage{bm}
\usepackage{epsfig}
\begin{document}
\title{Transient State Work Fluctuation Theorem for a Driven Classical Dissipative System }
\author{Rajarshi Chakrabarti}

\begin{abstract}
We derive the nonequilibrium transient state work fluctuation
theorem and also the Jarzynski equality for a classical harmonic
oscillator linearly coupled to a harmonic heat bath, which is
dragged by an external agent. Coupling with the bath makes the
dynamics not only dissipative but also non-Markovian in general.
Since we do not assume anything about the spectral nature of the
harmonic bath the derivation is not only restricted to the Markovian
bath rather it is more general, for a non-Markovian bath.

\end{abstract}
 \affiliation{Department of Inorganic and Physical chemistry,
Indian Institute of Science, Bangalore 560012, India} \maketitle

\section{Introduction}

Although the field of equilibrium thermodynamics and equilibrium
statistical mechanics are well explored, there existed almost no
theory for systems arbitrarily far from equilibrium until the advent
of fluctuation theorems (FTs)\cite{kurchan1, sevick1, ritort,
bustamante, evans1, evans2, gallavotti} in mid $90'$s. In general,
these fluctuation theorems have provided a general prescription on
energy exchanges that take place between a system and its
surroundings under general nonequilibrium conditions and explain how
macroscopic irreversibility appears naturally in systems that obey
time reversible microscopic dynamics. Fluctuation theorems have been
cast for various nonequilibrium quantities like heat, work, entropy
production etc and for systems obeying Hamiltonian \cite{dellago} as
well as stochastic dynamics \cite{vanzon, kurchan2}. Quantum
versions of FTs are also known \cite{wojcik, esposito}.

Apart from fluctuation theorems, Jarzynski \cite{jarzynski1,
jarzynski2} had also provided a remarkable relation between the work
done on a system with the equilibrium free energy difference. To
illustrate the relation, consider a classical system in contact with
a classical heat bath at temperature $T$. Initially the system was
in equilibrium with the bath and then driven by some external agent
(generalized force) $f$. Now the free energy $F$ for the system can
be calculated by computing the partition function $Z_{f}$ when the
generalized force is fixed at the value $f$, since
$F_f=-\beta^{-1}lnZ_f$, where $\beta=\frac1 {k_B T}$, $k_B$ is the
Boltzman constant. Let the system starts in equilibrium at time
$t=0$ specified by $f=A$ and then driven off to a later time
$t=\tau$. If this process is done quasistatically then the system
remains in equilibrium at each stages of the process and also at
$t=\tau$ specified by $f=B$. Then the work done $W$ on the system
equals the free energy difference, $\Delta F=F_{B}-F_{A}$. On the
other hand if the process is carried out with a finite rate (i.e. in
nonequilibrium conditions), $W$ will on average exceed $\Delta F$.

\begin{equation}
\label{inequal}\left\langle W\right\rangle \geq \Delta F
\end{equation}

The external agent $f$ is always varied in precisely the same manner
from $A$ to $B$. After each realization the work $W$ performed on
the system is calculated and the distribution function for $W$,
$P(W)$ is constructed. Jarzynski derived the following mathematical
equality popularly known as the Jarzynski equality (JE) or Jarzynski
relation where the angular bracket indicates the average taken over
the distribution function $P(W)$.

\begin{equation}
\label{jarzynski1}\left\langle \exp (-\beta W)\right\rangle =\exp
(-\beta \Delta F)
\end{equation}
\\

 Essentially there are two classes of fluctuation theorems, steady
state \cite{evans1, gallavotti} and transient fluctuation theorems
(TFT) \cite{evans2}. Since the paper deals with the transient state
fluctuation theorem for work here we do not discuss anything further
on steady state fluctuation theorem. The transient state work
fluctuation theorem gives the ratio of probabilities for the
production of positive work to the production of negative work as
follows,

\begin{equation}
\label{TFT}\frac{P(+W)}{P(-W)}=\exp (\beta W)
\end{equation}

Where, $W$ is the work done on the system by an external agent for
an arbitrary time period $\tau$. The theorem holds for any value of
$\tau$, provided one starts with the system in equilibrium.

Crooks \cite{crooks} provided another relation for the dissipative
work, $W_{diss}$, defined as $W_{diss}=W-\Delta F$. Where $\Delta F$
is the free energy difference between the final and the initial
equilibrium state. The relation is very similar to the transient
work fluctuation theorem and known as the Crooks fluctuation theorem
(CFT).

\begin{equation}
\label{crooks}\frac{P_F(+W_{diss})}{P_R(-W_{diss})}=\exp (\beta
W_{diss})
\end{equation}

Here $P_R(-W_{diss})$ is the probability distribution of negative
dissipative work done in a time-reversed process. Clearly $\Delta
F=0$ means $W=W_{diss}$. In such a situation the transient state
fluctuation theorem for work and the Crooks fluctuation theorem are
equivalent.

In this paper we derive the transient state fluctuation theorem for
work for a classical harmonic oscillator coupled linearly to a
harmonic bath. Because of the coupling to the bath, the system
becomes dissipative. We start from a Hamiltonian description for the
system plus the harmonic heat bath and then the system is driven by
an external agent for a time period of $\tau$ for a series of
measurements. We analytically calculate the distribution function
for work and show that it obeys the transient state fluctuation
theorem. For our particular choice of the external agent the free
energy change for the process is zero and hence the transient
fluctuation theorem and the Crooks fluctuation theorem are same.

In recent past, experiments \cite{wang1, wang2} were carried out to
verify the FTS. In the experiment by Wang \emph{et al.} a colloidal
particle was trapped using Laser and then dragged through a solvent
and subsequently the fluctuation theorem was verified. Our
Hamiltonian efficiently models such situation since the harmonic
oscillator in our model could be viewed as the colloidal particle in
the harmonic trap caused by Lasers and the harmonic bath as the
solvent through which the colloidal particle is dragged. People have
already verified FTs \cite{vanzon, astumian, saha} in the context of
the above mentioned experiment. This modeling was based on a
Langevin description which was Markovian. Only very recently Mai
\emph{et.al}\cite{mai}, Speck et.al \cite{speck} and Ohkuma
\cite{ohkuma} \emph{et.al} have reported derivations based on
generalized Langevin equation taking care of the Non-Markovian
nature of the dynamics. But here we do not start from a Langevin
description rather we start with a Hamiltonian description. Since we
couple our system Hamiltonian with a set of bath oscillators
effectively it produces a noise acting on the system which depends
on bath variable and thus the dynamics becomes stochastic. Also our
derivation does not assume anything about the spectral nature of the
bath and the particle coupling. Hence the results are of very
general validity. The paper is arranged as follows. In the next
section we define work done on the system. In section III we
introduce our model. Section IV contains the detailed derivation of
the fluctuation theorem and in the last section we conclude our
results.
\section{Definition of heat}

According to Jarzynski \cite{jarzynski2}, the work done on the
system (described by the Hamiltonian $H_S$) by an external agent $f$
acting on the system from $t=0$ to $t=\tau$ is defined as

\begin{equation}
\label{work1}W=\int\limits_0^\tau \dot f\frac{\partial H_S}{\partial
f}dt
\end{equation}

We use this definition.

\section {Our Model}

We consider a classical particle of mass $m$ described by a
positional coordinate $x$ which is linearly coupled to a set of
harmonic oscillators, each of unit mass described by a positional
coordinate $q_i (i=1,2,...N)$ forming a harmonic heat bath, a model
for the surrounding solvent. In the experiment by Wang. \emph{et al}
the colloidal particle was dragged through the solvent by using a
Laser trap. The particle experiences a harmonic trap whose minima
moves in time. To model such a situation we assume that the particle
$x$ is in a harmonic well whose  minima is time dependent. We thus
introduce the Hamiltonian \cite{zwanzigbook, weissbook, zwanzig}

\begin{equation}
\label{hamiltonian} H=H_s+H_B+h_{int}
\end{equation}

where

$$H_S=\frac{p_x^2}{2m}+\frac
k2(x-\alpha (t))^2$$ and

$$H_B + h_{int}=\sum\limits_{i=1}^{N}\left(
\frac{p_i^2}2+\frac{\omega _i^2}2\left( q_i-\frac{c_i}{\omega
_i^2}x\right) ^2\right)$$

$H_S$ is the Hamiltonian for the system, $H_B$ is that for the
harmonic bath and $h_{int}$ represents the coupling of the system
with the bath. Here $p_x$ and $p_i$ are the momenta for the particle
and the $i$-th bath coordinate, $k$ is the force constant of the
optical trap and $\alpha(t)$ is the time dependent mean position of
the harmonic trap.

\section{Derivation of TFT for work and JE}

The time evolution of the system plus bath is governed by the
Hamiltonian $H$. The equations of motion for the system and the bath
oscillators are

\begin{equation}
\label{pxdot}\dot p_x=-\frac{\partial H}{\partial x}=-k(x-\alpha
(t))-\sum\limits_{i=1}^{N}\frac{c_i^2}{\omega _i^2}x+\sum%
\limits_{i=1}^{N}c_iq_i
\end{equation}

\begin{equation}
\label{pidot}\dot p_i=-\frac{\partial H}{\partial q_i}=-\omega
_i^2q_i+c_ix
\end{equation}
\begin{equation}
\label{xdot}\dot x=\frac{\partial H}{\partial p_x}=\frac{p_x}m
\end{equation}
\begin{equation}
\label{qidot}\dot q_i=\frac{\partial H}{\partial p_i}=p_i
\end{equation}

In our case the external agent $f$ is $\alpha(t)$ and hence the work
done on the system is

\begin{equation}
\label{work2}W=\int\limits_0^\tau \dot \alpha (t)\frac{\partial H_S}{%
\partial \alpha }dt=\int\limits_0^\tau -k\dot \alpha (t)(x(t)-\alpha
(t))dt.
\end{equation}

Thus within the harmonic Hamiltonian description the work done on
the system $W$ is linear in $x$. As our Hamiltonian is quadratic and
we shall assume equilibrium distributions for the initial
distribution for the initial conditions, $W$ would have a Gaussian
probability distribution \cite{mai} with a mean  $ \left\langle
W\right\rangle $ and the variance $ \sigma _W^2=\left\langle
W^2\right\rangle -\left\langle W\right\rangle ^2 $. So the
distribution function for $W$ is

\begin{equation}
\label{pw}P(W)=\frac 1{\sqrt{2\pi \sigma _W^2}}e^{-(W-\left\langle
W\right\rangle )^2/2\sigma _W^2}
\end{equation}

with

\begin{equation}
\label{waverage}\left\langle W\right\rangle =\int\limits_0^\tau
-k\dot \alpha (t)(\left\langle x(t)\right\rangle -\alpha (t))dt
\end{equation}

and

\begin{equation}
\label{variance}
 \sigma _W^2=k^2\int\limits_0^\tau
dt_1\int\limits_0^\tau dt_2\dot \alpha (t_1)\dot \alpha
(t_2)C(t_1,t_2)
\end{equation}

with $\Delta x(t_1)=x(t_1)- \left\langle x(t_1)\right\rangle$ and
$C(t_1,t_2)=\left\langle \Delta x(t_1)\Delta x(t_2)\right\rangle.$

With the above Gaussian distribution function for work it is easy to
show that $ \frac{P(W)}{P(-W)}=e^{\frac{2W\left\langle
W\right\rangle }{\sigma _W^2}} $. So in order to satisfy the TFT for
work it is enough to show that $ \sigma _W^2=\frac{2\left\langle
W\right\rangle }\beta $.

 Now it is obvious that to calculate the work done, its mean and the variance
 one has to know $x$ as a function of time $t$. To find $x$ as a
function of time $t$ we proceed as follows. We take a Laplace
transform of Eq. (\ref{pxdot})and Eq. (\ref{pidot})to get.

\begin{equation}
\label{xs2}\tilde x(s)=\frac{\left( k\tilde \alpha (s)+m\dot
x(0)+\sum\limits_{i=1}^{N}c_i\tilde q_i(s)+msx(0)\right) }{\left(
ms^2+k+\sum\limits_{i=1}^{N}\frac{c_i^2}{\omega _i^2}\right) }
\end{equation}

and

\begin{equation}
\label{qis2}\tilde q_i(s)=\frac{\left( p_i(0)+sq_i(0)+c_i\tilde
x(s)\right) }{\left( s^2+\omega _i^2\right) }
\end{equation}

Now we substitute $\tilde q_i(s)$ in Eq. (\ref{xs2}) from Eq.
(\ref{qis2}) and get $\tilde x(s)$ in terms of the initial momenta
and position of the bath coordinates in time and that of the system
itself.

\begin{equation}
\label{xsfinal}\tilde x(s)=k\tilde \alpha (s)+(p_x(0) + mx(0)s
+\tilde g(s))\tilde b(s)
\end{equation}

Where

\begin{equation}
\label{bs}\tilde b(s)=\frac 1{\left( k+ms^2+\sum\limits_{i=1}^{N}\frac{c_i^2%
}{\omega _i^2}-\sum\limits_{i=1}^{N}\frac{c_i^2}{(s^2+\omega
_i^2)}\right) }
\end{equation}

and

\begin{equation}
\label{gs}\tilde g(s)=\sum\limits_{i=1}^{N}c_i\frac{\left(
p_i(0)+sq_i(0)\right) }{\left( s^2+\omega _i^2\right) }
\end{equation}

 Taking the inverse Laplace transform of Eq. (\ref{xsfinal})
 one gets $x(t)$.

\begin{widetext}
\begin{equation}
\label{xtfinal}
\begin{array}{c}
x(t)=mx(0)y(t)+mv(0)b(t)  +\int\limits_0^tdt^{\prime }b(t-t^{\prime
})\left( k\alpha (t^{\prime })+\xi (t^{\prime })\right)
\end{array}
\end{equation}
\end{widetext}

with

$$ \xi (t)=g(t)-\sum\limits_{i=1}^{N}\frac{c_i^2}{\omega _i^2}\cos
(\omega _it)x(0) $$ and $$ y(t)=\int\limits_0^tdt^{\prime
}\left(\dot b(t^{\prime })\delta (t-t^{\prime })+\frac
1m\sum\limits_{i=1}^{N}\frac{c_i^2}{\omega _i^2}\cos (\omega
_it^{\prime })b(t-t^{\prime })\right) $$

Now Eq. (\ref{xtfinal}) should be consistent with the initial
conditions. This readily gives

 \begin{equation}
\label{my0}my(0)=1
\end{equation}
\begin{equation}
\label{mb0}mb(0)=0
\end{equation}
\begin{equation}
\label{mb0dot}m\dot b(0)=1
\end{equation}
\begin{equation}
\label{my0dot}m\dot y(0)=0
\end{equation}

One can substitute $x(t)$ from Eq. (\ref{xtfinal}) to Eq.
(\ref{work2}) to calculate $W$. Next task is to calculate $
\left\langle W\right\rangle $ which is obtained from Eq.
(\ref{work2}) by replacing $x(t)$  with its thermal average

\begin{widetext}
\begin{equation}
\label{xtaveragefinal}
\begin{array}{c}
\left\langle x(t)\right\rangle =m\left\langle x(0)\right\rangle
y(t)+m\left\langle v(0)\right\rangle b(t)+\int\limits_0^tdt^{\prime
}b(t-t^{\prime })\left( k\alpha (t^{\prime })+\left\langle \xi
(t^{\prime })\right\rangle \right)
\end{array}
\end{equation}
\end{widetext}

 Here the angular bracket indicates a thermal average taken in the initial state of the harmonic bath (at
$t=0$) with the shifted canonical equilibrium distribution (since we
start from an initial equilibrium distribution) given by $ \rho \sim
e^{-\beta H^{(0)}} $. Thus $ \left\langle p_x(0)\right\rangle =0$, $
\left\langle x(0)\right\rangle =\alpha (0)$. Now to calculate $
\left\langle g(t) \right\rangle$ we proceed as follows. First we
take the inverse Laplace transform of Eq. (\ref{gs}) to get

\begin{widetext}
\begin{equation}
\label{gt}
\begin{array}{c}
g(t)
=\sum\limits_{i=1}^{N}c_i\left\{ p_i(0)\frac{\sin (\omega _it)}{\omega _i}%
+q_i(0)\cos (\omega _it)\right\}
\end{array}
\end{equation}
\end{widetext}

Next we take the thermal average with respect to the initial
distribution $ \rho \sim e^{-\beta H^{(0)}} $ to get

\begin{equation}
\label{gtaverage}
\begin{array}{c}
\left\langle g(t)\right\rangle
=\sum\limits_{i=1}^{N}\frac{c_i^2}{\omega _i^2}x(0)\cos (\omega _it)
\end{array}
\end{equation}

as $ \left\langle p_i(0)\right\rangle =0 $, $ \left\langle
q_i(0)\right\rangle =\frac{c_i}{\omega _i^2}x(0) $.  Thus
$\left\langle \xi (t)\right\rangle =\left\langle g(t)\right\rangle
-\sum\limits_{i=1}^{N}\frac{c_i^2}{\omega _i^2}\cos (\omega
_it)x(0)=0$. The quantity $\xi(t)$ is a Gaussian random force from
the bath with the statistical properties, $ \left\langle \xi
(t)\right\rangle =0 $, $ \left\langle \xi (t)\xi (t^\prime
)\right\rangle =\beta ^{-1}\Gamma (t-t^{\prime }) $, where
$\Gamma(t)=\sum\limits_{i=1}^{N}\frac{c_i^2}{\omega _i^2}\cos
(\omega _it) $.

Finally we get

\begin{equation}
\label{xtaveragefinal2}\left\langle x(t)\right\rangle =m\alpha
(0)y(t)+\int\limits_0^tdt^{\prime }k\alpha (t^{\prime
})b(t-t^{\prime })
\end{equation}

Next we derive a set of equations in the Laplace and the time
domain. These are used to get the TFT for work.

 In a few steps one can show, $m s
\tilde y(s)=1-k \tilde b(s)$, where $y(s)=s \tilde b(s)+ \frac
{\tilde \Gamma (s)}{m} \tilde b(s)$, $\tilde \Gamma
(s)=\sum\limits_{i=1}^{N}\frac{sc_i^2}{\omega _i^2(s^2+\omega _i^2)}
$. Inverse Laplace transform of $m s \tilde y(s)=1-k \tilde b(s)$
gives $\dot y(t)=-(\frac k m)b(t)$. Also $s \tilde b(s)=\tilde
y(s)-\frac {\tilde \Gamma(s)}{m} \tilde b(s)$ whose inverse Laplace
gives $ \dot b(t)=y(t)-\frac 1m\int\limits_0^tdt^{\prime }\Gamma
(t^{\prime })b(t-t^{\prime }) $. So we have the following set of
important relations in the Laplace and the time domain.

\begin{equation}
\label{relations}
\begin{array}{c}
ms\tilde y(s)=1-k\tilde b(s) \\
\\
s\tilde b(s)=\tilde y(s)-\frac 1m\tilde \Gamma (s)\tilde b(s) \\
\\
\dot y(t)=-\left( \frac km\right) b(t) \\
\\
\dot b(t)=y(t)-\frac 1m\int\limits_0^tdt^{\prime }\Gamma (t^{\prime
})b(t-t^{\prime })
\end{array}
\end{equation}

Using the above set of equations one can show that the average work
done on the system (which is given by Eq. (\ref{waverage})) is

 \begin{equation}
\label{waveragefinal}\left\langle W\right\rangle =\left( \frac
mk\right) \int\limits_0^\tau dt\int\limits_0^tdt^{\prime }k\dot
\alpha (t)y(t-t^{\prime })k\dot \alpha (t^{\prime })
\end{equation}

To derive it we proceed as follows. First we replace $\left\langle
x(t)\right\rangle$ in Eq. (\ref{waverage}) from
Eq.(\ref{xtaveragefinal2}) to get

$$
\begin{array}{c}
\left\langle W\right\rangle =\frac k2\left(
\frac{\alpha ^2(\tau)}2-\frac{\alpha ^2(0)}2\right)  \\  \\
-\int\limits_0^\tau dtkm\alpha (0)\dot \alpha
(t)y(t)-\int\limits_0^\tau dt\int\limits_0^tdt^{\prime }k\dot \alpha
(t)b(t-t^{\prime })k\alpha (t^{\prime })
\end{array}
$$
and then using Eq. (\ref{relations}) followed by an integration by
parts and using Eq. (\ref{my0}) one ultimately gets

$$
\left\langle W\right\rangle =\left( \frac mk\right)
\int\limits_0^\tau dt\int\limits_0^tdt^{\prime }k\dot \alpha
(t)y(t-t^{\prime })k\dot \alpha (t^{\prime })
$$

which is Eq. (\ref{waveragefinal}).

In order to evaluate the variance we first have to calculate the
correlation function,  $C(t_1,t_2)=\left\langle \Delta x(t_1)\Delta
x(t_2)\right\rangle$. Now using Eq. (\ref{xtfinal}) one can show

\begin{widetext}
\begin{equation}
\label{c12l}
\begin{array}{c}
C(t_1,t_2) =m^2\left( k\beta \right) ^{-1}y(t_1)y(t_2)+m^2\left(
m\beta \right) ^{-1}b(t_1)b(t_2) +\beta
^{-1}\int\limits_0^{t_1}dt_1^{\prime
}\int\limits_0^{t_2}dt_2^{\prime }b(t_1-t_1^{\prime
})b(t_2-t_2^{\prime })\Gamma (t_1^{\prime }-t_2^{\prime })
\end{array}
\end{equation}
\end{widetext}

Where we have used $ \left\langle \xi (t)\xi (t^\prime
)\right\rangle =\beta ^{-1}\Gamma (t-t^{\prime }) $, $ \left\langle
\left( x(0)-\alpha (0)\right) ^2\right\rangle =\left( k\beta \right)
^{-1}\text{, }\left\langle v^2(0)\right\rangle =\left( m\beta
\right) ^{-1} $.

 The above expression for $C(t_1,t_2)$ is then put back in Eq. (\ref{variance}) to get

 \begin{widetext}
 \begin{equation}
 \label{variance2}
\begin{array}{c}
\sigma _W^2=\left( \frac{m^2}{\beta k}\right) \left(
\int\limits_0^\tau dtk\dot \alpha (t)y(t)\right) ^2 +\left( \frac
m\beta \right) \left( \int\limits_0^\tau dtk\dot \alpha
(t)b(t)\right) ^2 + \\
\frac 1\beta\int\limits_0^\tau
dt_1\int\limits_0^\tau dt_2\int\limits_0^{t_1}dt_1^{\prime
}\int\limits_0^{t_2}dt_2^{\prime } \Gamma (t_1^{\prime }-t_2^{\prime
})k\dot \alpha (t_1)b(t_1-t_1^{\prime })b(t_2-t_2^{\prime })k\dot
\alpha (t_2)
\end{array}
\end{equation}
\end{widetext}

Let us define

$$
J(t_1,t_2)=\int\limits_0^{t_1}dt_1^{\prime
}\int\limits_0^{t_2}dt_2^{\prime }\Gamma (t_1^{\prime }-t_2^{\prime
})b(t_1-t_1^{\prime })b(t_2-t_2^{\prime })
$$
 With the help of Laplace and Fourier transform one can show (see
 Appendix for the evaluation of the integral)

\begin{equation}
\label{jt1t2}
 J(t_1,t_2)=\left( \frac mk\right) y(t_1-t_2)-\left(
\frac {m^2}{k}\right) y(t_1)y(t_2)-mb(t_1)b(t_2)
\end{equation}

When the above expression for $J(t_1,t_2)$ is plugged into Eq.
(\ref{variance2}) one gets

$$
\sigma _W^2=\left( \frac m{k \beta}\right) \int\limits_0^\tau
dt_1\int\limits_0^\tau dt_2k\dot \alpha (t_1)y(t_1-t_2)k\dot \alpha
(t_2)
$$

In Eq. (\ref{waveragefinal}). $t$ and $t^{\prime}$ being dummy
variables one can change these into $t_1$ and $t_2$ respectively and
rewrite Eq. (\ref{waveragefinal}).

\begin{equation}
\label{waverage1}\left\langle W\right\rangle =\left( \frac mk\right)
\int\limits_0^\tau dt_1\int\limits_0^{t_1}dt_2k\dot \alpha
(t_1)y(t_1-t_2)k\dot \alpha (t_2)
\end{equation}

Now  we interchange  $t_1$ and $t_2$.

\begin{equation}
\label{waverage2}\left\langle W\right\rangle =\left( \frac mk\right)
\int\limits_0^\tau dt_2\int\limits_0^{t_2}dt_1k\dot \alpha
(t_2)y(t_2-t_1)k\dot \alpha (t_1)
\end{equation}

Adding Eq. (\ref{waverage1}). and Eq. (\ref{waverage2}). Since
 $y(t)$ is an even function of $t$  and also  $\tau >t_1$,  $\tau
>t_2$ one gets

$$
2\left\langle W\right\rangle =\left( \frac mk\right)
\int\limits_0^\tau dt_1\int\limits_0^\tau dt_2k\dot \alpha
(t_1)y(t_1-t_2)k\dot \alpha (t_2)=\beta \sigma _W^2
$$

This shows that the TFT for work is satisfied. Next we come to
Crooks fluctuation theorem. As we have mentioned earlier that TFT
for work and CFT becomes equivalent when the free energy change for
the concerned process becomes zero ($\Delta F=0$, $W=W_{diss}$)
which happens in our case. This is because we assumed the external
agent is the time dependent mean position of the harmonic trap and
subsequently when one evaluates the partition function by performing
a Gaussian integral, $\alpha (t)$ does not appear in the partition
function and thus the free energy becomes independent of $\alpha(t)$
so the free energy change is zero. Hence the TFT for work reads as

\begin{equation}
\label{tftfinal}\frac{P(+W)}{P(-W)}=\exp (\beta W)
\end{equation}

JE is obtained easily by integrating above equation

$$
\begin{array}{c}
\left\langle \exp (-\beta W)\right\rangle =\int\limits_{-\infty
}^\infty
dW P(+W)\exp (-\beta W) \\
=\int\limits_{-\infty }^\infty dW P(-W)=1
\end{array}
$$

\section{conclusions}

The paper verifies the TFT for work and the JE for a classical
dissipative system which is dragged through by an external agent. We
start from a Hamiltonian description of our system which is linearly
coupled to a bath. The coupling makes the dynamics stochastic and
non-Markovian in general. A non-Markovian bath is more realistic
because of the existence of finite correlation time of the noise
acting on the particle. As far as our knowledge goes this is the
first detailed derivation of the TFT for work for such classical
dissipative system starting from a Hamiltonian description rather
than from a Langevin equation. To keep our derivation analytic we
had to restrict ourselves to a harmonic system and a harmonic bath.

\section{acknowledgements}

The author is grateful to Prof. K. L. Sebastian for his useful
comments and encouragements. The author also acknowledges Council of
Scientific and Industrial Research (CSIR), India for financial
support. The author is also thankful to Prof. Abhishek Dhar for
arranging the summer school on fluctuation theorems.

\section{Appendix}

The integral $J(t_1,t_2)$ is evaluated as follows. First let us
consider the integral

\begin{widetext}
\begin{equation}
\label{eq1appen}
\begin{array}{c}
\int\limits_0^\infty dt_2e^{-s_2t_2}\int\limits_0^{t_2}dt_2^{\prime
}b(t_2-t_2^{\prime })e^{-i\omega t_2^{\prime }}
=b(s_2)\int\limits_0^\infty dt_2e^{-s_2t_2}e^{-i\omega t_2} =b(s_2)
\frac{e^{-(s_2+i\omega )t_2}}{(s_2+i\omega )}\mid _0^\infty
=\frac{b(s_2)}{(s_2+i\omega )}
\end{array}
\end{equation}
\end{widetext}

Similarly

\begin{equation}
\label{eq2appen}\int\limits_0^\infty
dt_1e^{-s_1t_1}\int\limits_0^{t_1}dt_1^{\prime }b(t_1-t_1^{\prime
})e^{i\omega t_1^{\prime }}=\frac{b(s_1)}{(s_1-i\omega )}
\end{equation}

Then with the help of Eq. (\ref{eq1appen}) and Eq. (\ref{eq2appen})
double Laplace transform of $J(t_1,t_2)$  can be written as

\begin{widetext}
\begin{equation}
\label{eq3appen}\tilde J(s_1,s_2)=\int\limits_0^\infty
dt_1e^{-s_1t_1}\int\limits_0^\infty
dt_2e^{-s_2t_2}\int\limits_0^{t_1}dt_1^{\prime
}\int\limits_0^{t_2}dt_2^{\prime }\Gamma (t_1^{\prime }-t_2^{\prime
})b(t_1-t_1^{\prime })b(t_2-t_2^{\prime })
\end{equation}
\end{widetext}

We also define

$\Gamma (t_1^{\prime }-t_2^{\prime })=\frac 1{\sqrt{2\pi }%
}\int\limits_{-\infty }^\infty d\omega \tilde \Gamma (\omega
)e^{i\omega (t_1^{\prime }-t_2^{\prime })}$ and

$\tilde \Gamma (\omega )=\frac 1{\sqrt{2\pi }%
}\int\limits_{-\infty }^\infty dt\Gamma (t)e^{-i\omega t}
$

Then after some algebraic manipulations  Eq. (\ref{eq3appen})
becomes

\begin{widetext}
\begin{equation}
\label{eq6appen}
\begin{array}{c}
\tilde J(s_1,s_2)=\int\limits_0^\infty
dt_1e^{-s_1t_1}\int\limits_0^\infty
dt_2e^{-s_2t_2}\int\limits_0^{t_1}dt_1^{\prime
}\int\limits_0^{t_2}dt_2^{\prime }b(t_1-t_1^{\prime
})b(t_2-t_2^{\prime })\left( \frac 1{ \sqrt{2\pi
}}\int\limits_{-\infty }^\infty d\omega \tilde \Gamma (\omega
)e^{i\omega (t_1^{\prime }-t_2^{\prime })}\right) \\ \\ =\frac
1{\sqrt{2\pi }}\int\limits_{-\infty }^\infty d\omega \frac{\tilde
b(s_1)\tilde b(s_2)\tilde \Gamma (\omega )}{(s_1-i\omega
)(s_2+i\omega )}=\frac{\tilde b(s_1)\tilde b(s_2)}{(s_1+s_2)%
\sqrt{2\pi }}\int\limits_{-\infty }^\infty d\omega \left(
\frac{\tilde
\Gamma (\omega )}{s_1-i\omega }+\frac{\tilde \Gamma (\omega )}{s_2+i\omega }%
\right)
\end{array}
\end{equation}
\end{widetext}

\begin{equation}
\label{eq7appen}\tilde J(s_1,s_2)=\frac{\tilde b(s_1)\tilde b(s_2)}{(s_1+s_2)%
\sqrt{2\pi }}\int\limits_{-\infty }^\infty d\omega \left(
\frac{\tilde
\Gamma (\omega )}{s_1-i\omega }+\frac{\tilde \Gamma (\omega )}{s_2+i\omega }%
\right)
\end{equation}

Next consider the integral $ \frac 1{\sqrt{2\pi
}}\int\limits_{-\infty }^\infty d\omega \frac{\tilde \Gamma (\omega
)}{s_1-i\omega } $ which can be evaluated as follows

\begin{widetext}
\begin{equation}
\label{eq8appen}
\begin{array}{c}
\frac 1{ \sqrt{2\pi }}\int\limits_{-\infty }^\infty d\omega
\frac{\tilde \Gamma
(\omega )}{s_1-i\omega }=\frac 1{\sqrt{2\pi }}\frac 1{\sqrt{2\pi }%
}\int\limits_{-\infty }^\infty \frac{d\omega }{(s_1-i\omega )}%
\int\limits_{-\infty }^\infty dte^{-i\omega t}\Gamma (t)
=\int\limits_{-\infty }^\infty dt\Gamma (t)\left( \frac 1{-2\pi
i}\int\limits_{-\infty }^\infty \frac{d\omega }{(\omega
+is_1)}e^{-i\omega t}\right) =\\ \\ \int\limits_{-\infty }^\infty
dt\Gamma (t)\Theta (t)e^{-s_1t}=\int\limits_0^\infty dt\Gamma
(t)e^{-s_1t}=\tilde \Gamma (s_1)
\end{array}
\end{equation}
\end{widetext}

Similarly

\begin{equation}
\label{eq9appen}\frac 1{\sqrt{2\pi }}\int\limits_{-\infty }^\infty
d\omega \frac{\tilde \Gamma (\omega )}{s_2+i\omega }=\tilde \Gamma
(s_2)
\end{equation}

Now we use Eq. (\ref{eq8appen}) and Eq. (\ref{eq9appen}) to get

\begin{equation}
\label{eq10appen}\tilde J(s_1,s_2)=\frac{\tilde b(s_1)+\tilde b(s_2)}{%
(s_1+s_2)}\left( \tilde \Gamma (s_1)+\tilde \Gamma (s_2)\right)
\end{equation}

The above Eq. is then simplified with the help of the Eq.
(\ref{relations}) to get

\begin{equation}
\label{eq13appen}\tilde J(s_1,s_2)=\left( \frac mk\right)
y(t_1-t_2)-\left( \frac{m^2}k\right) y(t_1)y(t_2)-mb(t_1)b(t_2)
\end{equation}

Taking the double Laplace transformation of the above Eq. gives

$$J(t_1,t_2)=\left( \frac mk\right) y(t_1-t_2)-\left(
\frac {m^2}{k}\right) y(t_1)y(t_2)-mb(t_1)b(t_2)$$

which is Eq. (\ref{jt1t2}).

\bibliographystyle{apsrev}


\end{document}